\documentclass[a4paper,12pt]{article}
\usepackage[english]{babel}
\usepackage{fullpage}
\usepackage{epsfig}
\usepackage{amsfonts}
\usepackage{amssymb}
\usepackage{amsmath}
\usepackage{theorem}

\numberwithin{equation}{section}

{  \theorembodyfont{\normalfont}
 }
{  \theorembodyfont{\normalfont}
 }
\makeatletter
\title{Symmetric invariant manifolds \\ in the Fermi-Pasta-Ulam lattice}
\author{Bob Rink\thanks{Mathematics Institute, Utrecht University,
  PO Box 80.010, 3508 TA Utrecht, The Netherlands,
Telephone: 003130-2534557, Fax: 003130-2518394, E-mail:
  {\tt rink@math.uu.nl}}}
\begin{document}
\maketitle 
\hyphenation{ma-ni-fold}
\abstract{\noindent The Fermi-Pasta-Ulam (FPU) lattice with periodic boundary conditions and $n$ particles admits a large group of discrete symmetries. The fixed point sets of these symmetries naturally form invariant symplectic manifolds that are investigated in this short note. For each $k$ dividing $n$ we find $k$ degree of freedom invariant manifolds. They represent short wavelength solutions composed of $k$ Fourier-modes and can be interpreted as embedded lattices with periodic boundary conditions and only $k$ particles. Inside these invariant manifolds other invariant structures and exact solutions are found which represent for instance periodic and quasi-periodic solutions and standing and traveling waves. Some of these results have been found previously by other authors via a study of mode coupling coefficients and recently also by investigating `bushes of normal modes'. The method of this paper is similar to the latter method and much more systematic than the former. We arrive at previously unknown results without any difficult computations. It is shown moreover that similar invariant manifolds exist also in the Klein-Gordon lattice and in the thermodynamic and continuum limits. }

\section{Introduction}
The Fermi-Pasta-Ulam (FPU) lattice is a discrete model for a continuous nonlinear string, introduced by E. Fermi, J. Pasta and S. Ulam \cite{Alamos}. This string is modeled by a finite number of point masses which represent the material elements of the string. Each of the point masses is an oscillator that interacts with its nearest neighbors only.\\ 

\begin{center}
\epsfig{file=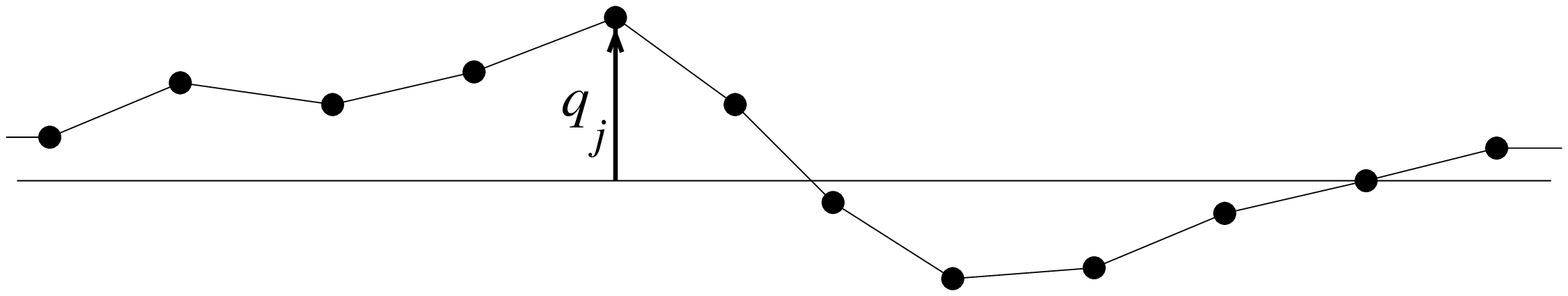, width=11cm, height=2cm}
\end{center}

\noindent Assume that the lattice consists of a finite number $n \in \mathbb{N}$ particles. Define $q_j \in \mathbb{R}$ the vertical position of the $j$-th
particle. We distinguish two different types of boundary conditions. We speak of fixed boundary conditions if the first and the last particle do not move, meaning that we have $q_0 = q_n = 0$ for all time. The FPU lattice with fixed boundary conditions models a string with Dirichlet boundary conditions. It is also possible to choose periodic boundary conditions, in which case the first and the last particle are identified, that is $q_0 = q_n$ for all time. The FPU lattice with periodic boundary conditions models a circular string. Both types of boundary conditions occur very often in the literature. In this paper we shall only consider lattices with periodic boundary conditions, as it will become clear that each lattice with fixed boundary conditions is naturally embedded as an invariant manifold of an appropriate periodic lattice. The particles of the periodic lattice are labeled by elements of the cyclic group $\mathbb{Z}/_{n
\mathbb{Z}}$. The Hamiltonian equations of motion for the FPU lattice are derived as follows.\\
\indent The space of positions $q=(q_1, \ldots, q_n)$ of the
particles in the lattice is $\mathbb{R}^n$. The space of positions and
conjugate momenta is the cotangent bundle $T^*\mathbb{R}^n$ of
$\mathbb{R}^n$, the elements of which are denoted $(q, p) = (q_1,
\ldots, q_n, p_1, \ldots, p_n)$. $T^*\mathbb{R}^n$ is a symplectic
manifold, endowed with the symplectic form $dq \wedge dp =
\sum_{j=1}^n dq_j \wedge dp_j$. Any smooth function $H: T^*\mathbb{R}^n
\to \mathbb{R}$ now induces the Hamiltonian vector field $X_H$ given by
the defining relation $(dq\wedge dp)(X_H, \cdot) = dH$. In other words,
we have the system of ordinary differential equations $\dot q_j =
\frac{\partial H}{\partial p_j}, \ \dot p_j = - \frac{\partial
H}{\partial q_j}$. \\
\indent The Hamiltonian function for the FPU lattice with periodic boundary conditions and $n$ particles consists of a kinetic energy and a potential energy. The potential energy is assumed to depend only on the vertical distance between pairs of neighboring particles. Hence the Hamiltonian is
\begin{equation}\label{hamfpu} H = \sum_{j \in \mathbb{Z}/_{n
\mathbb{Z}}} \frac{1}{2} p_j^2 + W(q_{j+1} - q_j) \ , \end{equation} in which $W: \mathbb{R} \to \mathbb{R}$ is a Lennard-Jones potential energy density function of
the form 
\begin{equation}\label{potentiaal}
W(x) = \frac{1}{2!} x^2 + \frac{\alpha}{3!} x^3 + \frac{\beta}{4!} x^4
 \ \ . 
\end{equation}
The $\alpha, \beta$ are real parameters measuring the
nonlinearity in the forces between the particles in the lattice. We also write 
$$H = \sum_{j \in \mathbb{Z}/_{n \mathbb{Z}}} \left( \frac{1}{2} p_j^2  + \frac{1}{2} (q_{j+1}-q_j)^2 \right) + \alpha H_3(q) + \beta H_4(q)\ .$$
In which 
$$H_m(q) = \frac{1}{m!} \sum_{j \in \mathbb{Z}/_{n \mathbb{Z}}} (q_{j+1} - q_j)^m$$
is a polynomial in $q$ of degree $m$. \\  
\indent For $\alpha = \beta = 0$, the Hamiltonian is quadratic and the equations of motion linear. The solutions can be written down explicitly and the motion is in fact completely integrable. But for $\alpha, \beta \neq 0$, the system is much harder to analyse. It can be interpreted as a nonlinear perturbation of the integrable linear Hamiltonian system. Fermi, Pasta and Ulam expected that a many particle system such as the FPU lattice would be ergodic due to these nonlinearities, meaning that almost all orbits densely
fill up an energy level set of the Hamiltonian $H$. Ergodicity would in theory
have to lead to `thermalisation', i.e. equipartition of energy between the 
various Fourier modes of the system.  FPU's nowadays famous numerical experiment was
intended to investigate how and at what time-scale this thermalisation would take place. The
result was astonishing: it turned out that there was no sign of
thermalisation at all. Putting initially all the energy 
in one Fourier mode, they observed that this energy was shared by only
a few other modes, the remaining modes were hardly
excited. Within a rather short time the system returned
close to its initial state and thus behaved more or less quasi-periodically.\\
\indent The observations of Fermi, Pasta and Ulam greatly stimulated work on nonlinear dynamical systems. Nowadays people tend to explain the FPU experiment in two ways. In
1965 Zabuski and Kruskal \cite{Kruskal} considered the Korteweg-de Vries 
equation as a continuum limit of the FPU lattice and numerically
found the first indications for the stable behaviour of solitary
waves. We now know that the Korteweg-de Vries equation is integrable. This clearly suggests an explanation for FPU's observations, although our understanding of the relation between the FPU lattice and its infinite dimensional limits has until now been quite disappointing.\\
\indent  Another, possibly correct explanation for the
quasi-periodic behaviour of the FPU system, is based on the
Kolmogorov-Arnol'd-Moser (KAM) 
theorem. As is well-known
\cite{arnold}, the solutions of 
an $n$ degree of freedom Liouville
integrable Hamiltonian system are constrained to move on 
$n$-dimensional tori
 and are not at all ergodic but periodic and quasi-periodic.
The KAM theorem states that most invariant
tori of such an integrable system persist under small Hamiltonian
perturbations, if the unperturbed integrable system satisfies a certain nondegeneracy condition which is called the Kolmogorov condition. This shows that quasi-periodic behaviour can also be observed in certain nonintegrable Hamiltonian systems. Although several authors,
starting with Izrailev and Chirikov 
\cite{Chirikov}, have
therefore stated that the KAM theorem explains the observations of the FPU
experiment, it has for a long time been completely unclear how the FPU system can
be viewed as a
perturbation of a nondegenerate integrable system. At least the linear Hamiltonian system for the FPU lattice with $\alpha=\beta=0$ does certainly not satisfy the Kolmogorov condition. This gap in the theory was recently mentioned again in the review article of Ford \cite{Ford} and the book of Weissert \cite{Weissert}. \\
\indent The only rigorous results in this direction that are known to me were obtained by Nishida \cite{Nishida} and Rink \cite{Rink2}. These authors compute a Birkhoff normal form for the FPU lattice. This Birkhoff normal form is an approximation of the Hamiltonian of the FPU lattice. Although the results in \cite{Nishida} are unfortunately incomplete, it is proven in \cite{Rink2} that this Birkhoff normal form is indeed often nondegenerately integrable. This explains why the FPU lattice at low energy can not be ergodic but must display a lot of quasi-periodic behaviour. On the other hand, many numerical studies indicate that above a certain energy threshold the lattice indeed thermalises. Reference \cite{poggiruffo} contains a rather complete overview of these results. \\ 
\indent Contrary to these more or less global approximation results, several authors have been trying to find exact low dimensional invariant manifolds for the FPU lattice. First of all because they represent interesting classes of solutions such as periodic and quasi-periodic solutions and standing and traveling waves. But also because it is believed by some authors, see for instance \cite{Budinsky}, that the destabilisation of invariant manifolds can lead to chaos and hence maybe to ergodicity. The present paper is inspired by the idea that the results of \cite{Rink2}, which were primarily obtained for FPU lattices with periodic boundary conditions, will to a large extent also be applicable to subsystems of these periodic lattices. We will see for instance that every FPU lattice with fixed boundary conditions can be viewed as such a subsystem. This paper does not contain any explicit KAM statements though, but focusses on finding invariant manifolds only. \\
\indent Most of the invariant manifolds that are known in the FPU lattice were discovered more or less emperically. In their original paper Fermi, Pasta and Ulam \cite{Alamos} already remarked that if the nonlinearity coefficient $\alpha$ in (\ref{potentiaal}) vanishes and initially only waves with an odd wave number are excited, then waves with an even wave number will never gain energy. Later on, other invariant manifolds were discovered by studying mode coupling coefficients in detail, see for instance \cite{Bivins} and \cite{poggiruffo}. In these papers it is shown that certain sets of normal modes will not be excited if they initially have no energy. \\
\indent As will be explained in section \ref{q-p}, studying mode coupling coefficients can be quite unsatisfactory. A more systematic method for finding invariant manifolds in a physical system should be based on the symmetries of this system. The only reference that exploits these symmetries for the FPU lattice is \cite{chechinfpu} in which so-called `bushes of normal modes' are computed. These `bushes' are simply invariant manifolds of a certain type. Their definition and how to find them are discussed more elaborately in \cite{chechin}. The basic idea is the well-known physical principle that the fixed point set of a symmetry forms an invariant manifold for the equations of motion. In \cite{chechinfpu}, several previously unknown  `bushes' are classified. After computing the irreducible representations of the symmetry group of the FPU lattice and introducing appropriate `symmetry-adapted coordinates', the computation of these `bushes of normal modes' is fairly simple. \\
\indent In the present paper it will be shown that the previously mentioned invariant manifolds and many others can be found even without introducing Fourier modes and studying mode coupling coefficients and without computing irreducible representations and symmetry-adapted coordinates. We only have to compute the fixed point sets of the various symmetries. As we incorporate more symmetries then \cite{chechinfpu} in our considerations, we find various invariant manifolds that were not discussed before, in particular for the so-called $\beta$-lattice. Moreover, our results are not only valid for the FPU lattice, but for any lattice with the same symmetries, such as the Klein-Gordon lattice \cite{Morgante}. They also apply in the thermodynamic limit as the number of particles grows large, and in the continuum limit: we can point out several infinite dimensional invariant manifolds for a rather broad class of nonlinear homogeneous partial differential equations. Some of them have been found previously in a very unpractical way.

\section{Quasi-particles}\label{q-p}
Since we want to be able to compare our results with previous work, we introduce Fourier modes in this section. These Fourier modes are at the same time the `symmetry-adapted coordinates' of \cite{chechinfpu}. It is natural to view the solutions of the FPU lattice as a superposition of waves and to make the following Fourier transformation:
\begin{align}
&q_j = \frac{1}{\sqrt{n}} \sum_{k \in \mathbb{Z}/_{n\mathbb{Z}}} e^{\frac{2 \pi i j k }{n}} \bar q_k \\
&p_j =  \frac{1}{\sqrt{n}} \sum_{k \in \mathbb{Z}/_{n\mathbb{Z}}} e^{-\frac{2 \pi i j k }{n}} \bar p_k
\end{align}
Using that $$\frac{1}{n}\sum_{k \in \mathbb{Z}/_{n\mathbb{Z}}} e^{\frac{2 \pi i j k}{n}} = \left\{ \begin{array}{ll} 1 & \mbox{if} \ j = 0\ \mbox{mod}\ n\\ 0 & \mbox{if}\ j \neq 0\ \mbox{mod}\ n \end{array} \right.$$
one easily calculates that $\{\bar q_j , \bar q_k\} = \{\bar p_j , \bar p_k\} = 0$ and $\{\bar q_j , \bar p_k\} = \delta_{jk}$, the Kronecker delta. Hence, $(\bar q, \bar p)$ are canonical coordinates. They are traditionally called {\it phonons} or {\it quasi-particles}. Written out in phonons, the FPU Hamiltonian (\ref{hamfpu}) reads as follows. The kinetic energy becomes:
$$\sum_{j \in \mathbb{Z}/_{n \mathbb{Z}}} \frac{1}{2} p_j^2 =  \frac{1}{2}\bar p_{n}^2 + \frac{1}{2}\bar p_{\frac{n}{2}}^2 + \sum_{1 \leq j < \frac{n}{2}} \bar p_j \bar p_{n-j} \ ,$$
where it is understood that the term $\frac{1}{2}\bar p_{\frac{n}{2}}^2$ occurs only if $n$ is even. The potential energies $H_m$ (for $m=2,3,4$) become
\begin{align} \nonumber
H_m &= \frac{1}{m!} \sum_{j \in \mathbb{Z}/_{n \mathbb{Z}}} (q_{j+1} - q_j)^m\\ \nonumber &= \frac{1}{m!} \sum_{j \in \mathbb{Z}/_{n \mathbb{Z}}} \left( \frac{1}{\sqrt{n}} \sum_{k \in \mathbb{Z}/_{n\mathbb{Z}}} (e^{\frac{2\pi i (j+1) k }{n}} - e^{\frac{2\pi i j k }{n}})\ \bar q_k \right) ^m \\ \nonumber &=  
\frac{1}{m!n^{\frac{m}{2}}}  
%\sum_{j \in \mathbb{Z}/_{n \mathbb{Z}}}  \sum_{\theta: |\theta|=m}
\hskip-.2cm \sum_{\begin{array}{c} \scriptscriptstyle  j \in \mathbb{Z}/_{n \mathbb{Z}} \\ \scriptscriptstyle \theta: |\theta|=m \end{array}} \hskip-.2cm
\left( \hskip-.15cm \begin{array}{c} m  \\  \theta  \end{array}\hskip-.15cm \right) e^{\frac{2 \pi i j ({\Large \Sigma}_k k \theta_k)}{n}} \prod_{k \in \mathbb{Z}/_{n \mathbb{Z}}}  (e^{\frac{2 \pi i k}{n}} - 1)^{\theta_k} \ \bar q_k^{\ \theta_k} \\ \nonumber &= n^{\frac{2-m}{2}}\hskip-.6cm \sum_{\begin{array}{c} \scriptscriptstyle  \theta: |\theta|=m \\ \scriptscriptstyle \Sigma_k k \theta_k = 0\!\!\!\!\! \mod n \end{array}} \hskip-.5cm  \prod_{k \in \mathbb{Z}/_{n \mathbb{Z}}} \frac{1}{\theta_k!}(e^{\frac{2 \pi i k}{n}} - 1)^{\theta_k}\ \bar q_k^{\ \theta_k}
\end{align}\noindent in which the sum is taken over multi-indices $\theta \in \mathbb{Z}^n$ for which $|\theta| := \sum_k |\theta_k| = m$. We also used the multinomial coefficient $(^m_{\hspace{.06cm} \theta}) := \frac{m!}{\Pi_k \theta_k!}$. We have obtained a rather compact and tractible formula for the Hamiltonian in phonon-coordinates. \\ 
\indent Let us also introduce real-valued phonons. For $1 \leq k < \frac{n}{2}$ define {\footnotesize
$$Q_k \! = \! (\bar q_k + \bar q_{n-k})/\sqrt{2} = \sqrt{\frac{2}{n}} \hspace{-.2cm} \sum_{j \in \mathbb{Z}/_{n \mathbb{Z}}} \hspace{-.2cm}  \cos(\frac{2jk\pi}{n}) q_j \ , \ Q_{n-k} \! = \! i(\bar q_k - \bar q_{n-k})/\sqrt{2} = \sqrt{\frac{2}{n}} \hspace{-.2cm}  \sum_{j \in \mathbb{Z}/_{n \mathbb{Z}}} \hspace{-.2cm} \sin(\frac{2jk\pi}{n}) q_j $$
$$P_k \! = \! (\bar p_k + \bar p_{n-k})/\sqrt{2} = \sqrt{\frac{2}{n}} \hspace{-.2cm} \sum_{j \in \mathbb{Z}/_{n \mathbb{Z}}} \hspace{-.2cm} \cos(\frac{2jk\pi}{n}) p_j \ , \ P_{n-k} \! = \! i(\bar p_{n-k} - \bar p_{k})/\sqrt{2} = \sqrt{\frac{2}{n}} \hspace{-.2cm} \sum_{j \in \mathbb{Z}/_{n \mathbb{Z}}} \hspace{-.2cm} \sin(\frac{2jk\pi}{n}) p_j $$}
\noindent and {\footnotesize $$Q_{\frac{n}{2}} = \bar q_{\frac{n}{2}} = \frac{1}{\sqrt{n}}\sum_{j=1}^n(-1)^j q_j\ , \ P_{\frac{n}{2}} = \bar p_{\frac{n}{2}} = \frac{1}{\sqrt{n}}\sum_{j=1}^n(-1)^j p_j\ ,$$ $$ \ Q_n = \bar q_n = \frac{1}{\sqrt{n}}\sum_{j=1}^n q_j \ , \ P_n = \bar p_n = \frac{1}{\sqrt{n}}\sum_{j=1}^n p_j\ .$$}\\
\noindent The transformation $(q, p) \mapsto (Q, P)$ is again symplectic and one can express the Hamiltonian in terms of $Q$ and $P$. In the case that $\alpha = \beta =0$, that is for the harmonic FPU lattice, one gets 
$$H = \sum_{j \in \mathbb{Z}/_{n \mathbb{Z}}} \frac{1}{2} p_j^2 + \frac{1}{2} (q_{j+1} - q_j)^2 = \sum_{j = 1}^{n}\frac{1}{2}(P_j^2 + \omega_j^2 Q_j^2) $$
in which for $j=1, \ldots, n$ the numbers $\omega_j$ are the well-known normal mode frequencies of the periodic FPU lattice: $$\omega_j := 2 \sin (\frac{j \pi}{n})$$
Note that written down in real-valued phonon coordinates, the equations of motion of the harmonic lattice are simply the equations for $n-1$ uncoupled harmonic oscillators and one free particle. The situation is not so simple anymore if $\alpha, \beta \neq 0$, when the normal modes interact in a complicated manner that is governed by the Hamiltonians
$$H_m = \sum_{\theta:|\theta|=m} c_{\theta} \prod_{k = 1}^{n-1} Q_k^{\theta_k} $$
in which the $c_{\theta}$ are certain coefficients. An expression for the $c_{\theta}$ can in principle be obtained from the formulas for the Hamiltonian $H_m(\bar q)$ and the mapping $\bar q \mapsto Q$. For instance $H_4(Q)$ can explicitly be found in \cite{poggiruffo}, although its computation is not given there. \\
\indent Note that $H$ is independent of $Q_n=\bar q_n = \frac{1}{\sqrt{n}}\sum_j q_j$. Hence the total momentum $P_n=\bar p_n = \frac{1}{\sqrt{n}}\sum_j p_j$ is a constant of motion and the equations for the remaining variables are completely independent of $(Q_n,P_n)=(\bar p_n, \bar q_n)$. It is common to set the latter coordinates equal to zero, or to neglect them completely. Thus one removes the total momentum from the equations of motion. Equivalently, one could also perform the Marsden-Weinstein reduction of the symmetry induced by the flow of $X_{\bar p_n} = \frac{\partial}{\partial \bar q_n}$, cf. \cite{A&M} or \cite{Rink2}.  \\
\indent In the nonlinear system, the other normal modes interact in a complicated manner, even though not every possible coupling term occurs. Only those monomials $\bar q^{\theta} = \prod_k \bar q_k ^{\theta_k}$ are present in $H_m(\bar q)$ for which $\sum_k k \theta_k = 0 \! \! \mod \! n$, whereas $H_m(Q)$ contains only the monomials $Q^{\theta} = \prod_k Q_k ^{\theta_k}$ for which $c_{\theta} \neq 0$. In the next section we will see that this is a consequence of discrete symmetries in the system. \\
\indent It is exactly the fact that not every coupling term occurs which accounts for the existence of various invariant manifolds, see \cite{Bivins} and \cite{poggiruffo}. Let $\mathcal{A} \subset \mathbb{Z}/_{n\mathbb{Z}}$. Then the manifold spanned by modes in $\mathcal{A}$ is
$$M_n^\mathcal{A} := \{(Q,P)\in T^*\mathbb{R}^n|Q_j = P_j = 0\ \forall  j \notin \mathcal{A} \}\ .$$
In several cases, these $M_n^\mathcal{A}$ are invariant manifolds for the equations of motion. In \cite{chechin} and \cite{chechinfpu} they are then called `bushes of normal modes'. We will not use this terminology. One readily infers from the equations of motion $\dot Q_j =
\frac{\partial H}{\partial P_j}, \ \dot P_j = - \frac{\partial
H}{\partial Q_j}$ that $M_n^\mathcal{A}$ is an invariant manifold (a `bush') if and only if $c_{\theta}=0$ for all $\theta$ with the property that $\theta_j = 1$ for some $j \notin \mathcal{A}$ and $\theta_k = 0$ for all $k \notin \mathcal{A} \cup \{j\}$. Making use of this fact, several invariant manifolds have been discovered. If $n$ is even, one can for instance choose $\mathcal{A} = \{\frac{n}{2}\}$. It is then obvious that $\mathcal{A}$ satisfied the required property since $j + (m-1)\frac{n}{2} \neq 0 \mod n$. The solutions in the invariant manifold $M_n^{\{\frac{n}{2}\}}$ are of the form $q_j(t) = \frac{(-1)^j}{\sqrt{n}}Q_{\frac{n}{2}}(t)$. This type of periodic solutions in which neighbouring particles are exactly out of phase, is well-known. In \cite{poggiruffo} a linear stability analysis is given for this solution in the $\beta$-lattice (i.e. $\alpha =0$) and in \cite{chechinfpu} a similar linear stability analysis is given for this solution in the $\alpha$-lattice (i.e. $\beta = 0$).\\
\indent Studying mode coupling coefficients in this way, several invariant manifolds have been discovered. In \cite{Bivins} it is shown that if $\alpha = 0$ and $n$ is even, $M_n^{\{2,4,\ldots,n\} }$ and  $M_n^{\{1,3,\ldots,n-1\} }$ are invariant. Poggi and Ruffo \cite{poggiruffo} show that $M_n^{\{\frac{n}{3}, \frac{2n}{3}\}}$ and $M_n^{\{\frac{n}{4}, \frac{3n}{4}\}}$ are invariant.
\\
\indent The above method is rather simple and easily understood but has the following limitations:
\begin{itemize}
\item[1.] An explicit expression for the $c_{\theta}$ is required.
\item[2.] The method becomes more elaborate if one wants to find invariant manifolds of higher dimensions. 
\item[3.] There is no a priori `physical' reason why a certain $M_n^\mathcal{A}$ will be invariant.
\item[4.] Invariant manifolds might exist that are not of the form $M_n^\mathcal{A}$ for some $\mathcal{A} \subset \mathbb{Z}/_{n\mathbb{Z}}$. 
\item[5.] It is not clear whether the discovered invariant manifolds will also  be present in the continuum limit or in other one-dimensional lattice systems.
\end{itemize}
For these reasons, studying mode coupling coefficients is rather unsatisfactory. With the method presented in the following sections of this paper it is possible to detect easily many more invariant manifolds. They arise in a natural way as fixed point sets of symmetries.

\section{Symmetry}
The Hamiltonian function (\ref{hamfpu}) of the periodic FPU lattice has discrete symmetries with important dynamical consequences. Let us discuss symmetries in general here. Assume that $P:T^*\mathbb{R}^n \to T^*\mathbb{R}^n$ is a linear isomorphism with the following two properties:
\begin{itemize}
\item[1.] $P$ is symplectic, i.e. $P^*(dq\wedge dp) = dq \wedge dp$.
\item[2.] $P$ leaves the Hamiltonian invariant, i.e. $P^*H:= H \circ P = H$. 
\end{itemize}
Under these assumptions, $P$ is called a symmetry of $H$. The set of symmetries of $H$ is a group under composition of functions. This group is denoted $G_H$.\\
\indent For every symmetry $P\in G$ we find that the Hamiltonian vector field $X_H$ induced by $H$ is
equivariant under $P$: $P^*X_H = X_{P^*H} = X_H$. In other words:
if $\gamma:\mathbb{R} \to T^*\mathbb{R}^{n}$ is an integral curve of
$X_H$, then $P \circ \gamma: \mathbb{R} \to T^*\mathbb{R}^{n}$ is also an
integral curve of $X_H$. This implies that $P$ commutes with the flow of $X_H$, that is $e^{tX_H} \circ P = P \circ e^{tX_H}$. \\
\indent Of particular dynamical interest is the fixed point set of a symmetry $P$,
\begin{equation}
\mbox{Fix} \ P = \{ (q, p) \in T^*\mathbb{R}^n | P(q,p) = (q,p) \} 
\end{equation}
Let $(q,p) \in \mbox{Fix}\ P$, then $P(e^{tX_H} (q,p)) =e^{tX_H} (P(q,p)) =  e^{tX_H} (q,p)$. So $\mbox{Fix} \ P$ is an invariant manifold for the flow of $X_H$. This explains why fixed point sets are so interesting. When $G' \subset G_H$ is a subgroup, then a fixed point set is also defined for it: $\mbox{Fix}\ G' = \cap_{P\in G'} \mbox{Fix}\ P$. These are of course also invariant manifolds and they are commonly studied.  \\
\indent Fixed point sets of symmetries and fixed point sets of subgroups have a very simple relation. When $P_1, \ldots, P_m$ are symmetries, then $\langle P_1, \ldots, P_m \rangle \subset G_H$ is the smallest subgroup of $G_H$ containing $P_1, \ldots, P_m$. The symmetries $P_1, \ldots, P_m$ are called generators for this subgroup. One readily checks now that $\mbox{Fix}\ \langle P_1, \ldots, P_m \rangle  = \cap_j \mbox{Fix}\ P_j$. It therefore suffices to study the fixed point sets of seperate symmetries. If $G'$ is a subgroup of $G$ that is generated by the symmetries $P_1, \ldots, P_m$, then the fixed point set of $G'$ is simply the intersection of the fixed point sets of the seperate symmetries $P_1, \ldots, P_m$. As the number of elements of $G_H$ can be considerably less then the number of subgroups of $G_H$, we prefer to study fixed point sets of seperate symmetries first and take their intersections later. \\
\\
\noindent Let us discuss the symmetries of the FPU lattice now. Define the linear mappings $R, S, T:
\mathbb{R}^{n} \to \mathbb{R}^{n}$ by
\begin{align}\label{RSactie}
&R: (q_1, q_2, \ldots, q_{n-1}, q_n) \mapsto (q_2, q_3, \ldots, q_n, q_1)  \\ \nonumber
&S: (q_1, q_2, \ldots, q_{n-1}, q_n) \mapsto (-q_{n-1}, -q_{n-2}, \ldots, -q_{1}, -q_{n}) \\
\nonumber
&T: (q_1, q_2, \ldots, q_{n-1}, q_n) \mapsto (-q_1, -q_2, \ldots, -q_{n-1}, -q_n)  
\end{align}
The mappings $(q, p) \mapsto (Rq, Rp)$, $(q, p) \mapsto (Sq, Sp)$ and $(q, p) \mapsto (Tq, Tp)$ from $T^*\mathbb{R}^n$ to $T^*\mathbb{R}^n$ are also denoted $R$, $S$ and $T$ respectively. They satisfy the multiplication relations $R^n = S^2 = T^2 = \mbox{Id}$ and $RS = SR^{-1}$, while $T$ commutes with everything. Hence the discrete group $\langle R,S \rangle$ $ :=$ $\{\mbox{Id},$
$R, R^2,$ $ \ldots, R^{n-1},$ $S, RS,$ $\ldots, R^{n-1}S\}$ is a representation of the $n$-th dihedral group $D_n$, the symmetry group of the $n$-gon, whereas 
$\langle R, S, T \rangle$ is a representation of $D_n \times \mathbb{Z}/_{2 \mathbb{Z}}$. \\
\indent $R$, $S$ and $T$ are symplectic maps and $R$ and $S$
leave the Hamiltonian $H$ invariant. $T$ only leaves $H$ invariant if the potential energy density function $W$ is an even function, in other words if $\alpha=0$. When $\alpha \neq 0$, then $G_H = \langle R, S \rangle$ is the symmetry group of $H$, whereas $G_H=\langle R, S, T \rangle$ if $W$ is even.\\
\\
\noindent In the coming sections we shall investigate the various invariant manifolds $\mbox{Fix}\ P$ for $P \in G_H$. We shall describe them in terms of the original coordinates $(q,p)$, but also in phonon-coordinates $(\bar q, \bar p)$ and $(Q, P)$. Therefore it is interesting to write down how $R, S$ and $T$ act in complex phonon coordinates:
\begin{align}\label{RSactiephonon}
\nonumber
R: &(\bar q_1, \bar q_2, \ldots, \bar q_{n-1}, \bar q_n) \mapsto (e^{2\pi i / n}\bar q_1, e^{4\pi i / n}\bar q_2, \ldots, e^{2\pi i (n-1)/ n}\bar q_{n-1}, e^{2\pi i n / n}\bar q_n) \ , \\ \nonumber
&(\bar p_1, \bar p_2, \ldots, \bar p_{n-1}, \bar p_n) \mapsto (e^{-2\pi i / n}\bar p_1, e^{-4\pi i / n}\bar p_2, \ldots, e^{-2\pi i (n-1)/ n}\bar p_{n-1}, e^{-2\pi i n / n}\bar p_n) \ . \\ \nonumber
S: &(\bar q_1, \bar q_2, \ldots, \bar q_{n-1}, \bar q_n) \mapsto (-\bar q_{n-1}, -\bar q_{n-2}, \ldots, -\bar q_{1}, -\bar q_{n}) \ , \\ \nonumber 
&(\bar p_1, \bar p_2, \ldots, \bar p_{n-1}, \bar p_n) \mapsto (-\bar p_{n-1}, -\bar p_{n-2}, \ldots, -\bar p_{1}, -\bar p_{n})\ .\\
\nonumber
T: &(\bar q_1, \bar q_2, \ldots, \bar q_{n-1}, \bar q_n) \mapsto (-\bar q_1, -\bar q_2, \ldots, -\bar q_{n-1}, -\bar q_n) \ , \\ \nonumber
&(\bar p_1, \bar p_2, \ldots, \bar p_{n-1}, \bar p_n) \mapsto (-\bar p_1, -\bar p_2, \ldots, -\bar p_{n-1}, -\bar p_n) \ . 
\end{align}
Note that by performing the transformation to complex phonons, $R$ has been diagonalised, whereas the actions of $S$ and $T$ have not at all changed. This means that $(\bar q, \bar p)$ are what in \cite{chechinfpu} is called `symmetry-adapted coordinates'. They are actually adapted to the subgroup $\langle R \rangle$ of $G_H$.  The action of $R$ on a monomial $q^{\theta}$ is also very simple:
$$R^*\left(\prod_k q_k^{\theta_k}\right) = e^{2\pi i \sum_k k \theta_k / n} \ \prod_k q_k^{\theta_k} $$
In other words, the monomial $q^{\theta}$ is $R$-symmetric if and only if $\sum_k k \theta_k = 0 \! \! \! \mod \! n$. So $R$-symmetry is the reason why only these monomials occur in the FPU Hamiltonian.

\section{Invariant manifolds for arbitrary potentials}
\noindent In this section we study the invariant manifolds that are formed by the fixed point sets of elements of $G_H=\langle R, S \rangle \cong D_n$. So it is not yet assumed that the potential energy density function $W$ is even.\\
\indent For integers $n$ and $k$, let $\gcd(n,k)$ be the greatest common divisor of $n$ and $k$. For $k \in \mathbb{Z}$, 
$$\mbox{Fix} \ R^k = \{ q_j = q_{j+\gcd (n,k)}, p_j = p_{j+\gcd (n,k)} \ \forall j \}$$  is an invariant $\gcd(n,k)$ degree of freedom symplectic submanifold of $T^*\mathbb{R}^n$. The Hamiltonian function $H|_{ \mbox{Fix} \ R^k }$ on the symplectic submanifold $\mbox{Fix} \ R^k $ obviously simply models the periodic FPU lattice with $\gcd(n,k)$ particles. In this way, the periodic lattice with $k$ particles is naturally embedded in the lattice with $n$ particles if $k$ divides $n$. In phonon coordinates, 
\begin{align}\nonumber
\mbox{Fix} \ R^k =& \{\bar q_j = \bar p_j = 0 \ \forall j \neq 0 \!\!\!\! \mod \! \frac{n}{\gcd(n,k)} \}\\ \nonumber = \{Q_j &= P_j = 0 \ \forall j \neq 0 \!\!\!\! \mod \! \frac{n}{\gcd(n,k)} \}
\end{align} 
So if $k$ divides $n$, then $\mbox{Fix} \ R^k = M_n^{\{\frac{n}{k}, \frac{2n}{k}, \ldots, \frac{(k-1)n}{k}, n\}}$ and is hence spanned by modes which represent a repeating spatial pattern with period $k$.\\
\indent If for instance $n$ is even, then $\mbox{Fix} \ R^2 = M_n^{\{\frac{n}{2}, n\}}$ is the two degree of freedom invariant manifold spanned by the $\frac{n}{2}$-th and the $n$-th normal modes. If we as usual neglect the $n$-th mode, which moves independently of all other modes, we find that $\mbox{Fix}\ R^2$ consists of all solutions of the form $q_j(t) = \frac{(-1)^j}{\sqrt{n}} Q_{\frac{n}{2}}(t)$. These are the previously mentioned periodic solutions in which neighboring particles are exactly out of phase. On the other hand one has for even $n$ that $\mbox{Fix} \ R^{\frac{n}{2}} = M_n^{\{2,4,\ldots, n\}}$. It consists of all even modes. \\
\indent If $3$ divides $n$, then $\mbox{Fix}\ R^3 = M_n^{\{\frac{n}{3}, \frac{2n}{3}, n\}}$, whereas $\mbox{Fix}\ R^{\frac{n}{3}}=M_n^{\{3, 6, \ldots, n-3, n\}}$. Etcetera. These invariant manifolds were discussed already extensively in \cite{chechinfpu}.\\
\\
\noindent The following invariant manifolds are only briefly discussed in \cite{chechinfpu}. For arbitrary $l \in \mathbb{Z}$ we can study 
{\footnotesize
\begin{align} \nonumber
& \mbox{Fix} \ R^lS = \{q_j = -q_{l-j}, p_j = -p_{l-j} \ \forall j\} = \{\bar q_j = -e^{-\frac{2\pi i j l}{n}} \bar q_{n-j}\  , \ \bar p_j = -e^{\frac{2\pi i j l}{n}} \bar p_{n-j} \ \forall j\} = \\ \nonumber & \{ Q_j \cos(\frac{l j \pi}{n}) + Q_{n-j}\sin(\frac{l j\pi}{n}) = P_j \cos(\frac{l j \pi}{n}) + P_{n-j} \sin(\frac{l j\pi}{n}) = 0 \ \forall \ 1 \leq j < \frac{n}{2}  \ , \\ \nonumber & Q_{\frac{n}{2}} = (-1)^{l+1}Q_{\frac{n}{2}}\ , \ P_{\frac{n}{2}} = (-1)^{l+1}P_{\frac{n}{2}} \ , \ Q_n = P_n = 0 \}
\end{align} 
}
\noindent \!\!\!\! It is a $(2n-2-(-1)^{l}-(-1)^{n+l})/4$ degree of freedom symplectic subspace of $T^*\mathbb{R}^n$. \\
\indent Note that $\mbox{Fix}\ R^lS$ is not always of the form $M_n^\mathcal{A}$ for some $\mathcal{A}$. On the other hand, $\mbox{Fix}\ S = M_n^{\{j|\frac{n}{2} < j < n\}}$ and if $n$ is even, then $\mbox{Fix}\ R^{\frac{n}{2}} S = M_n^{\{1, n-2,3,n-4,\ldots\}}=M_n^{\{j|2\leq j\leq \frac{n}{2}, \ j = 1  \!\!\! \mod 2\}\cup\{j|\frac{n}{2}< j < n, \ j=0  \!\!\! \mod 2\}}$. So for instance for $n=8$ these are $M_8^{\{5,6,7\}}$ and $M_8^{\{1,3,6\}}$. \\
\indent If both $n$ and $l$ are even, then $\mbox{Fix} \ R^lS$ has dimension $n/2 -1$ and in $\mbox{Fix} \ R^lS$ we have $q_{\frac{l}{2}} = q_{\frac{n+l}{2}} = 0$. In other words, if $n$ is even, then for every even $l$ the Hamiltonian function $H|_{\mbox{Fix}\ R^lS}$ on the symplectic subspace $\mbox{Fix}\ R^lS$ models the FPU lattice with fixed boundary conditions and $n/2 - 1$ moving particles. Hence, the FPU lattice with fixed boundary conditions and $n/2 -1$ moving particles is naturally embedded in the periodic FPU lattice with $n$ particles. This is the reason why we do not study FPU lattices with fixed boundary conditions separately.

\section{Invariant manifolds for even potentials}
If the potential energy density function $W$ is even, then also $T$ is a symmetry and the full symmetry group of the FPU Hamiltonian is $G_H=\langle R, S, T \rangle \cong D_n \times \mathbb{Z}/_{2 \mathbb{Z}}$. Let us study the fixed point sets of the symmetries $R^kT$ and $R^lST$ which have not yet been discussed in the previous section. Most results in this section are new, as the symmetry $T$ was not considered in \cite{chechinfpu}.  \\
\indent For $k \in \mathbb{Z}$,
$$\mbox{Fix}\ R^kT = \{q_j = - q_{j+\gcd(n,k)}, p_j = -p_{j+\gcd(n,k)} \ \forall j\} $$
which is nontrivial only if $n/\gcd(n,k)$ is even -and hence $n$ must be even. In this case it is a $\gcd(n,k)$ degree of freedom invariant symplectic manifold. In phonons,
\begin{align}
\nonumber \mbox{Fix}\ R^kT =& \{ \bar q_j = \bar p_j = 0 \ \forall j \neq \frac{n}{2 \gcd(n,k)} \!\!\!\! \mod \! \frac{n}{\gcd(n,k)} \} \\ \nonumber = \{Q_j =& P_j = 0 \ \forall j  \neq \frac{n}{2\gcd(n,k)} \!\!\!\!  \mod  \! \frac{n}{\gcd(n,k)} \}\ .
\end{align} 
So if $2k$ divides $n$, then $\mbox{Fix}\ R^{k}T = M_n^{\{ \frac{n}{2k}, \frac{3n}{2k}, \ldots, \frac{(2k-1)n}{2k}\}}$. \\
\indent The special choice $k =\frac{n}{2}$ gives us the invariant manifold $\mbox{Fix}\ R^{\frac{n}{2}} T = M_n^{\{1,3,5,\ldots,n-1\}}$ of all odd normal modes that was already discovered by Fermi, Pasta and Ulam \cite{Alamos}. The choice $k = 1$ gives us $\mbox{Fix}\ R T = M_n^{\{\frac{n}{2}\}}$, the well known $\frac{n}{2}$-th mode. \\
\indent If $n$ is divisible by $4$, then $\mbox{Fix}\ R^{\frac{n}{4}} T = M_n^{\{2,6,10,\ldots,n-2\}}$ is invariant. This is a new result. The invariant manifold $\mbox{Fix}\ R^{2} T = M_n^{\{\frac{n}{4}, \frac{3n}{4}\}}$ is discussed in \cite{poggiruffo}. It contains quasi-periodic solutions.\\
\indent For an $n$ divisible by $6$ we find the invariant manifolds $M_n^{\{3, 9, 15, \ldots, n-3\}}$ and $M_n^{\{\frac{n}{6}, \frac{n}{2}, \frac{5n}{6}\}}$. Etcetera.\\
\\
\noindent For $l \in \mathbb{Z}$, 
{\footnotesize
\begin{align} \nonumber
& \mbox{Fix}\ R^l S T = \{q_j = q_{l-j}, p_j = p_{l-j} \ \forall j\} = \{\bar q_j = e^{-\frac{2\pi ij l}{n}} \bar q_{n-j} \ , \ \bar p_j = e^{\frac{2\pi ij l}{n}} \bar p_{n-j} \ \forall j\}  = \\ \nonumber & \{ Q_j \sin(\frac{l j \pi}{n}) - Q_{n-j}\cos(\frac{l j\pi}{n}) = P_j \sin(\frac{l j \pi}{n}) - P_{n-j} \cos(\frac{l j\pi}{n}) = 0 \ \forall \ 1 \leq j < \frac{n}{2} \ , \\ \nonumber & Q_{\frac{n}{2}} = (-1)^{l}Q_{\frac{n}{2}} \ , \ P_{\frac{n}{2}} = (-1)^{l}P_{\frac{n}{2}} \}
\end{align} 
}
\!\!\!\! is an $(2n-2+(-1)^{l}+(-1)^{n+l})/4$ degree of freedom invariant symplectic manifold. \\
\indent Note that again $\mbox{Fix}\ R^l S T$ is not always of the form $M_n^\mathcal{A}$, but that on the other hand $\mbox{Fix}\ ST = M_n^{\{j| 0 \leq j \leq \frac{n}{2}\}}$ and if $n$ is even, $\mbox{Fix}\ R^{\frac{n}{2}}ST = M_n^{\{0,n-1,2,n-3,4,\ldots\}}=M_n^{\{j|0\leq j\leq \frac{n}{2}, \ j=0  \!\!\! \mod 2\}\cup\{j|\frac{n}{2}< j < n, \ j=1 \!\!\! \mod 2\}}$. So for instance for $n=8$ these are $M_8^{\{1,2,3,4\}}$ and $M_8^{\{2,4,5,7\}}$.

\section{Examples of intersections}
We have studied the fixed point sets of the elements of the symmetry groups $G_H= \langle R, S\rangle$ and $G_H= \langle R, S, T \rangle$. They are equal to the fixed point sets of subgroups of $G_H$ that are generated by one element. A fixed point set of a subgroup generated by more than one element must be the intersection of some of the fixed point sets that were already discussed. We will give just a few examples here. \\
\\
\noindent If $3$ divides $n$, then $\mbox{Fix}\ R^3 \cap \mbox{Fix} \ S = M_n^{\{\frac{2n}{3}\}}$, whereas $\mbox{Fix}\ R^3 \cap \mbox{Fix} \ ST = M_n^{\{\frac{n}{3}\}}$. The latter is only invariant if the potential $W$ is even.\\
\indent If $4$ divides $n$, then $\mbox{Fix}\ R^4 \cap \mbox{Fix} \ S = M_n^{\{\frac{3n}{4}\}}$, $\mbox{Fix}\ R^4  \cap \mbox{Fix} \ ST = M_n^{\{\frac{n}{4}, \frac{n}{2}\}}$ and $\mbox{Fix}\ R^2T  \cap \mbox{Fix} \ ST = M_n^{\{\frac{n}{4}\}}$. \\
\indent If $5$ divides $n$, then $\mbox{Fix}\ R^5 \cap \mbox{Fix} \ S = M_n^{\{\frac{3n}{5},\frac{4n}{5}\}}$, whereas $\mbox{Fix}\ R^5  \cap \mbox{Fix} \ ST = M_n^{\{\frac{n}{5}, \frac{2n}{5}\}}$.\\
\indent If $6$ divides $n$, then $\mbox{Fix}\ R^6 \cap \mbox{Fix} \ S = M_n^{\{\frac{2n}{3},\frac{5n}{6}\}}$ and $\mbox{Fix}\ R^6  \cap \mbox{Fix} \ ST = M_n^{\{\frac{n}{6}, \frac{n}{3}, \frac{n}{2} \}}$. And we find that $\mbox{Fix} \ R^3T = M_n^{\{\frac{n}{6}, \frac{n}{2}, \frac{5n}{6}\}}$ can be split into $\mbox{Fix} \ R^3T \cap \mbox{Fix}\ S =M_n^{\{\frac{5n}{6}\}}$ and $\mbox{Fix} \ R^3T \cap \mbox{Fix}\ ST =M_n^{\{\frac{n}{6},\frac{n}{2}\}}$. The normal mode solutions for the $\beta$-lattice that lie in $M_n^{\{\frac{5n}{6}\}}$ have as far as I know never been discussed in the literature.\\
\\
\noindent One can proceed and compute, if $k$ divides $n$, the intersections of the various fixed point sets of $R^k, S, R^{\frac{n}{2}}S, R^kT, ST$ and $R^{\frac{n}{2}}ST$. We choose not to make a systematic classification of the results, since most invariant manifolds in the FPU lattice are not even of the form $M_n^{\mathcal{A}}$ for some $\mathcal{A}$.

\section{Other lattices and the continuum limit} 
A major advantage of our method is that fixed point sets of symmetries are invariant manifolds in any Hamiltonian system admitting these symmetries. Hence we expect to find the invariant manifolds that we discovered in the FPU lattice with periodic boundary conditions also in other one-dimensional spatially homogeneous lattices, such as the Klein-Gordon (KG) lattice \cite{Morgante}. The KG lattice with periodic boundary conditions has the Hamiltonian
$$H = \sum_{j \in \mathbb{Z}/_{n
\mathbb{Z}}} \frac{1}{2} p_j^2 + \frac{1}{2}(q_{j+1} - q_j)^2 + W(q_j) \ , $$
in which $W$ is a potential energy density function. The KG lattice models a one dimensional mono-atomic structure with small coupling between the atoms. It is clear that the mappings $R$ and $ST$, see formulas (\ref{RSactie}), again leave this Hamiltonian invariant, whereas $R$, $S$ and $T$ separately have this property if $W$ is an even function. Thus we have again found symmetries and their fixed point sets are invariant manifolds. In particular, the invariant manifolds that we discovered in the FPU lattice with even potential are also present in the KG lattice with even potential.\\
\\
\noindent  In the thermodynamic limit it is assumed that the FPU lattice consists of a countably infinite number of particles, labeled by $j\in \mathbb{Z}$. The equations of motion are Hamiltonian equations on $T^*\mathbb{R^Z}$ with Hamiltonian $$H = \sum_{j\in\mathbb{Z}} \frac{1}{2}p_j^2 + V(q_{j+1}-q_j)\ .$$
The symmetries are now induced by 
$$ \begin{array}{lll} 
R & : (\ldots, q_{-1}; q_0, q_1, \ldots) \mapsto (\ldots, q_{-1},q_0; q_1, \ldots) \\
S & : (\ldots, q_{-1}; q_0, q_1, \ldots) \mapsto (\ldots, -q_{1},-q_0; -q_{-1}, \ldots)\\
T & :(\ldots, q_{-1}; q_0, q_1, \ldots) \mapsto (\ldots, -q_{-1};-q_0, -q_{1}, \ldots) \end{array}$$
The finite dimensional manifold $\mbox{Fix}\ R^n$ models an infinite lattice with a spatially repeating pattern of period $n$. Or, equivalently, the periodic lattice with $n$ particles. Inside $\mbox{Fix}\ R^n$ we find again the invariant structures that were discussed previously in this paper. The invariant manifold $\mbox{Fix}\ R^nS$ is an infinite dimensional one. It consists of solutions with $q_j = - q_{n-j}$ that are anti-symmetric around $j=n/2$. Etcetera. Similar conclusions hold of course for the thermodynamic limit of the KG lattice.  \\
\\
\noindent Our results are also valid in the continuum limit, when the discrete lattice equations are replaced by a homogeneous partial differential equation. Consider for example for $x \in \mathbb{R}/_{\mathbb{Z}}$ the equation 
$$u_{tt} = u_{xx} + f(u)\ ,$$
for $f:\mathbb{R}\to \mathbb{R}$. This equation can also be written as the system of equations
$$u_t = v \ , \ v_t = u_{xx} + f(u)\ ,$$
which have the Hamiltonian 
$$H = \int_{\mathbb{R}/_{\mathbb{Z}}} \frac{1}{2}v(x)^2 + \frac{1}{2}u_x(x)^2 - F(u(x))\ dx$$
in which $F'=f$. Define the symplectic operators 
$$ \begin{array}{lll} \mathcal{R}^a & : u(\cdot) \mapsto u(a + \cdot), &  v(\cdot) \mapsto v(a + \cdot) \\
\mathcal{S} & : u(\cdot) \mapsto -u(-\ \cdot), &  v(\cdot) \mapsto -v(-\ \cdot)\\
\mathcal{T} & : u(\cdot) \mapsto -u(\cdot), & v(\cdot) \mapsto -v(\cdot) \ .\end{array}
$$
The constant $a \in \mathbb{R}/_{\mathbb{Z}}$ is arbitrary. Clearly, $H$ is invariant under $\mathcal{R}^a$ and $\mathcal{ST}$. $H$ is invariant under $\mathcal{R}^a$, $\mathcal{S}$ and $\mathcal{T}$ separately if and only if $F$ is even, that is if and only if $f$ is odd. \\
\indent The fixed point sets of these symmetries are invariant manifolds, possibly of infinite dimension. If $a \notin \mathbb{Q}$, then $\mbox{Fix}\ \mathcal{R}^a$ consists of constant solutions only, but if $a = \frac{p}{q}$ is rational and $\gcd(p,q)=1$, then $\mbox{Fix}\ \mathcal{R}^{\frac{p}{q}}$ represents the solutions with $u(t,x)=u(t,x+\frac{1}{q})$. $\mbox{Fix}\ \mathcal{R}^{\frac{1}{q}}\mathcal{T}$ consists of solutions with $u(x) = -u(x+\frac{1}{q})$. The latter is nontrivial only if $q$ is even. For arbitrary $a$, $\mbox{Fix}\ \mathcal{R}^a\mathcal{S}$ contains solutions with $u(x)=-u(a-x)$ and $\mbox{Fix}\ \mathcal{R}^a\mathcal{ST}$ represents solutions with $u(x)=u(a-x)$.\\
\indent It is natural to use the Fourier transformation
$$u(x,t) = \sum_{k\in \mathbb{Z}} u_k(t) e^{i k \pi x} \ , \ v(x,t) = \sum_{k\in \mathbb{Z}} v_k(t) e^{i k \pi x}$$  
and to express the fixed point sets in terms of the Fourier variables $(u_k, v_k)_{k\in \mathbb{Z}}$. We then find for instance the following invariant manifolds
$$\mbox{Fix}\ \mathcal{R}^{\frac{p}{q}} = \{ u_k = v_k = 0\ \forall k \neq 0 \!\!\! \mod q\} = M^{\{\ldots, -2q, -q, 0, q, 2q, \ldots\}}$$
$$\mbox{Fix}\ \mathcal{R}^{\frac{p}{q}}\mathcal{T} = \{ u_k = v_k = 0\ \forall k \neq q \!\!\! \mod 2q\} = M^{\{\ldots, -3q, -q, q, 3q, \ldots\}}$$
Etcetera.\\
\indent \cite{krol}, \cite{stroucken} and \cite{verhulst2} study the equation $u_{tt}=u_{xx} + u^3$ by the Galerkin-averaging method. By an analysis of mode coupling coefficients they discover that the manifolds $M^{\{\ldots, -2q, -q, 0, q, 2q, \ldots\}}$ and $M^{\{\ldots, -3q, -q, q, 3q, \ldots\}}$ are invariant in a certain finite dimensional system of differential equations, the Galerkin-averaging approximation, which approximates the original partial differential equation. We arrive here at the much stronger result that their conclusions hold for any odd nonlinearity $f$ and in the {\it original} partial differential equation.

\section{Discussion}
In a systematic way we found various invariant manifolds for the FPU oscillator lattice with periodic boundary conditions. These invariant manifolds represent interesting classes of solutions such as periodic and quasi-periodic solutions, standing and traveling waves and embedded lower dimensional FPU lattices with periodic or fixed boundary conditions. They are moreover interesting since it is believed by some authors \cite{Budinsky} that destabilisation of these invariant manifolds can lead to chaos. Some of the invariant structures that we found have previously been discovered by other authors by an analysis of mode coupling coefficients. Our method on the contrary is similar to the method of `bushes of normal modes' and looks for fixed point sets of symmetries which are natural invariant manifolds. We can derive our results without computing mode coupling coefficients explicitly. In fact, it is not even necessary to introduce normal modes at all as an expression for the invariant manifolds can simply be obtained in terms of the original physical variables, which are the positions and momenta of the particles in the lattice. In this way, we find several previously undiscussed invariant manifolds in the FPU lattice. The same invariant manifolds are present in other homogeneous Hamiltonian lattices such as the Klein-Gordon lattice and even in lattices with an infinity of particles. In the continuum limit, when the lattice equations are replaced by a homogeneous partial differential equation, we point out analogous infinite dimensional invariant structures.

\section{Acknowledgement}
The author would like to thank Giovanni Gallavotti, Dario Bambusi, Ferdinand Verhulst and one of the referees for many discussions and valuable hints.
\bibliography{/user1/home7/Rink/bibliografie/bibliografie}
\bibliographystyle{amsplain}

\end{document}